\documentclass[Journal,twocolumn,10pt]{IEEEtran}
\usepackage{graphicx}
\usepackage{amsfonts}
\usepackage{amssymb}
\usepackage{amsmath}
\usepackage{fancyhdr}
\usepackage{lastpage}
\usepackage{subfigure}

\usepackage{algorithm}
\usepackage{algpseudocode}
\usepackage{algorithmicx}

\newcommand{\comment}[1]{}

     \newenvironment{example}[1][Example]{\begin{trivlist}
     \item[\hskip \labelsep {\bfseries #1}]}{\end{trivlist}}

     \newcommand{\qed}{\nobreak \ifvmode \relax \else
           \ifdim\lastskip<1.5em \hskip-\lastskip
           \hskip1.5em plus0em minus0.5em \fi \nobreak
           \vrule height0.75em width0.5em depth0.25em\fi}
\begin{document}

% paper title

\title{Symbol-Index-Feedback Polar Coding Schemes for Low-Complexity Devices }

\author{\authorblockN{Xudong Ma \\}
\authorblockA{Pattern Technology Lab LLC, U.S.A.\\
Email: xma@ieee.org} }

\maketitle

\begin{abstract}

Recently, a new class of error-control codes, the polar codes, have
attracted much attention. The polar codes are the first known class
of capacity-achieving codes for many important communication
channels. In addition, polar codes have low-complexity encoding
algorithms. Therefore, these codes are favorable choices for
low-complexity devices, for example, in ubiquitous computing and
sensor networks. However, the polar codes fall short in terms of
finite-length error probabilities, compared with the
state-of-the-art codes, such as the low-density parity-check codes.
In this paper, in order to improve the error probabilities of the
polar codes, we propose novel interactive coding schemes using
receiver feedback based on polar codes. The proposed coding schemes
have very low computational complexities at the transmitter side. By
experimental results, we show that the proposed coding schemes
achieve significantly lower error probabilities.

\end{abstract}

\section{Introduction}

\label{section_introduction}

Recently, a new type of error control codes {\it polar codes} have
attracted much attention. Recently invented in 2009 \cite{arikan09},
these codes are the first known class of error control codes, which
achieve the Shannon channel capacity for binary input symmetric
output channels. In addition, the encoding algorithms of polar codes
have very low complexities compared with these of Turbo codes and
low-density parity-check codes. Thus, these codes are attractive
choices for low-complexity and power-constrained communication
devices. However, for finite-length performance, the polar codes
still fall short in terms of error probabilities compared with the
state-of-the-art error control codes, such as, the low-density
parity-check codes. Therefore, it is desirable that the error
probabilities of the polar codes can be improved.

It is well-known in the information theory that receiver feedback
can be used to improve the performance of channel coding. It is
shown by Schalkwijk and  Kailath that if the feedback channel is
noiseless, then doubly exponential error probabilities can be
achieved \cite{schalkwijk68}. It is shown by Wyner that if a peak
energy constraint is imposed, then only singly exponential error
probabilities are possible \cite{wyner68}.  Burnashev shows some
coding schemes using noiseless feedback and an upper bound on the
error probability exponent \cite{burnashev76}. The Burnashev bound
can be achieved by a scheme of Yamamoto and Itoh \cite{yamamoto79}.
Forney proposes a scheme based on decoding reliability estimation
and erasure feedback, which can achieve error probability exponents
strictly larger than the sphere-packing bounds \cite{forney68}. The
above schemes by Burnashev, Yamamoto-Itoh, and Forney are all based
on block-wise feedback. There also exist symbol-wise feedback
schemes, such as in \cite{horstein63} \cite{Sahai08} etc.

From these previous discussions, many important theoretical results
on achievable rates, error probabilities, reliable functions have
been obtained. However, there exist very few practical
low-complexity coding schemes using receiver feedback to achieve
these performance improvements. The existing symbol-wise feedback
schemes usually have high computational complexities. Some practical
Yamamoto-Ito type feedback schemes are discussed in \cite{caire06}.
The Yamamoto-Ito schemes are based on block-wise feedback and must
be built upon some baseline block codes, which can provide
reasonably reliable decoding results. The schemes in \cite{caire06}
use the Turbo and low-density parity-check codes as the baseline
block codes.

In this paper, we propose novel low-complexity coding schemes based
on polar codes using receiver feedback. We call these coding
schemes, {\it symbol-index-feedback} polar coding schemes. In the
proposed coding schemes, the data transmission process is mainly an
interactive process between the transmitter and the receiver. During
each time slot, the receiver determines the index for the next
transmit symbol and sends the index using the feedback channel.
After receiving the symbol index from the feedback channel, the
transmitter transmits the corresponding symbol to the receiver using
the forward channel. There exist at least two variations of our
proposed coding schemes, one fixed-length type and one
variable-length type. The differences are mainly on the stop rules
for the above interactive process. The next transmit symbol index
should be determined based on decoding reliability at the receiver
side. We propose a virtual equivalent channel approach for
determining the next transmit symbol index.

By experimental results, we show that the proposed coding schemes
achieve significantly lower error-probabilities compared with the
conventional polar codes. The symbol-index-feedback schemes can be
used as standalong symbol-wise feedback schemes. These schemes can
also be combined with the block-wise feedback schemes, such as the
Yamamoto-Ito type schemes. In the latter case, the
symbol-index-feedback schemes are used as baseline blocks codes. In
both the two cases, the proposed coding schemes have very low
computational complexities at the transmitter side.

The proposed coding schemes in this paper are attractive choices for
many low complexity devices in ubiquitous computing and sensor
network applications. In these applications, the transmitters and
receivers have rather different power and complexity constraints.
The transmitters, such as Radio Frequency IDentification (RFID) tags
and sensor nodes, usually have limited power supplies. For example,
these devices are usually battery based. The RFID tags and sensor
nodes must also have low-costs and low-complexities. On the other
hand, the receivers, such as the RFID readers and data collection
centers, usually are much less restricted in terms of costs and
power. There usually also exist reverse communication links from the
receiver to the transmitter, which can be used as feedback channels
in the proposed symbol-index-feedback polar coding schemes.

The rest of this paper is organized as follows. In Section
\ref{section_basic_polar}, we provide a basic review of the polar
codes. In Section \ref{section_equ_compound_channel}, we propose the
virtual equivalent channel method to determine decoding reliability.
The decoding reliability estimation may be used in the proposed
coding schemes to determine the next transmit symbol. In Section
\ref{section_polar_with_feedback}, we show the proposed
symbol-index-feedback polar coding schemes. Some numerical results
and discussions are shown in Section \ref{sec_numerical}. The
numerical results show that our schemes achieve significantly lower
error probabilities compared with the conventional polar codes. The
concluding remarks are present in Section \ref{sec_conclusion}.

We will use the following notation throughout this paper. We  use
$X_{n}^{N}$ to denote a sequence of symbols $(X_n, X_{n+1}, \ldots,
X_{N-1}, X_{N})$. We use $\oplus$ to denote the binary XOR operator.
For a binary input channel $X\rightarrow Y$, $X\in\{0,1\}$,
$y\in{\mathcal Y}$, the Bhattacharyya parameter $Z(Y)$ is defined as
\begin{align}
Z(Y)=\sum_{y\in{\mathcal Y}} \sqrt{{\mathbb P}(y|X=0){\mathbb
P}(y|X=1)}.
\end{align}

\section{Polar Codes}
\label{section_basic_polar}

In this section, we provide a brief review of polar codes. Polar
codes are a class of linear block codes with block length $N=2^L$,
where $L$ are certain positive integers. Each polar code is
associated with an index set ${\mathcal M}\subset \{1,2,\ldots,N\}$.
Let $K=|{\mathcal M}|$ the cardinality of the set $\mathcal M$. The
rate of the polar code is $K/N$.

The encoding process of a polar code includes two steps. In the
first step, the encoding method maps the binary string $M_{1}^{K}$
of the transmit message into one binary string $U_{1}^{N}$. Each bit
in $M_{1}^{K}$ is written into one of the $U_n$ with $n \in
{\mathcal M}$. For $n\notin {\mathcal M} $, $U_n=0$. In the second
step of encoding, the binary string $U_1^N$ is input into a so
called $W_N$ channel. The output of the $W_N$ channel is an $N$ bit
binary sequence $X_1^N$, which is the encoded codeword.

\begin{figure*}
 \centering
 \includegraphics[width=6in]{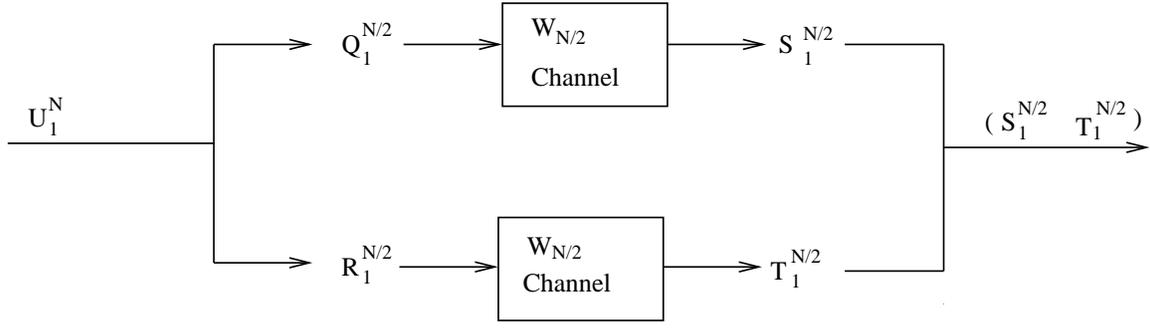}
 \caption{The $W_N$ channel}
 \label{wn_channel}
\end{figure*}

The $W_N$ channels are recursively defined. For a $W_2$ channel, if
the input is one binary sequence $(U_1,U_2)$, then the output of the
$W_2$ channel is the binary sequence $(U_1 \oplus U_2, U_2)$. For
one $W_N$ channel $N>2$, the channel first divides the input
$U_{1}^{N}$ into a binary odd-index-component substring $(U_1, U_3,
U_5, \cdots , U_{N-1})$ and a binary even-index-component substring
$(U_2, U_4, U_6, \cdots , U_{N})$. The even-index-component
substring is then added into the odd-index-component substring. That
is, two sequences are generated,
\begin{align}
& Q_{1}^{N/2}=(U_1\oplus U_2,
U_3\oplus U_4, \ldots, U_{N-1}\oplus U_N) \notag \\
& R_{1}^{N/2} = (U_2,U_4,\cdots,U_N) \notag
\end{align}
The sequences $Q_{1}^{N/2}$ and
$R_{1}^{N/2}$ are input into two independent $W_{N/2}$ channels. Let
$S_{1}^{N/2}$ and $T_{1}^{N/2}$ denote the outputs of the two
independent $W_{N/2}$ channels respectively. The output of the $W_N$
channel $X_{1}^{N}$ is the concatenated sequence of $S_{1}^{N/2}$
and $T_{1}^{N/2}$. A block diagram of the $W_N$ channel is shown in
Fig. \ref{wn_channel}.

\begin{example}
Suppose that we have a polar code with block length $4$. Suppose
that ${\mathcal M} = \{3,4\}$. Let the transmit message $M_1^2 =
(1,0)$. Then in the encoding process, $M_1^2$ is mapped into a
binary string $U_1^4 = (0,0,1,0)$. In the $W_4$ channel, $U_1^4$ is
decomposed into $Q_1^2 = (0,1)$ and $R_1^2 = (0,0)$. The outputs of
the two independent $W_2$ channels are $S_1^2 = (1,1)$, and $T_1^2 =
(0,0)$. The codeword $X_1^4 = (1,1,0,0)$.
\end{example}

The major decoding method of polar codes is the decision feedback
decoding algorithm.   The decoding algorithm sequentially estimates
the bits in the string $U_{1}^{N}$, starting from the first bit.  If
$n\notin {\mathcal M}$, then the decoding algorithm always decodes
$U_n = 0$. For each $n\in {\mathcal M}$, the algorithm makes the
decoding decision based on some calculated probabilities. Let
$Y_{1}^{N}$ denote the $N$ channel observations. The bit $U_n$ is
decoded into $0$, if
\begin{align}{\mathbb P}(U_n = 0| U_1^{n-1}, Y_{1}^{N})
\geq {\mathbb P}(U_n = 1|U_{1}^{n-1},Y_{1}^{N})
\label{eq_decoding_rule}
\end{align}
Otherwise, The bit $U_n$ is decoded into $1$. The sequence
$U_{1}^{n-1}$ can be taken to be the already decoded bits in the
probability calculation.

In \cite{arikan09}, the reliability of the decoding is analyzed by
using the Bhattacharyya parameters. Essentially for each $n\in
{\mathcal M}$, the bit-wise estimation can be related to a channel
\begin{align}
U_{n} \rightarrow \left(U_{1}^{n-1}, Y_{1}^{N}\right) \notag
\end{align}
For each such channel, we may define  a Bhattacharyya parameter
\begin{align}
Z(U_n) = \sum_{u_1^{n-1}, y_1^N} \sqrt{{\mathbb
P}(u_1^{n-1},y_1^N|0){\mathbb P}(u_1^{n-1},y_1^N|1)}
\end{align}
where, the summation is over all the realizations $u_1^{n-1},y_1^N$
of the random variables $U_1^{n-1},Y_1^N$.

It is shown in \cite{arikan09} that the Bhattacharyya parameters
$Z(U_n)$ can be recursively upper bounded. Note that $Q_1^{N/2}$ and
$R_1^{N/2}$ are the inputs of  some $W_{N/2}$ channels similarly as
$U_1^N$. Therefore,  similar channels and corresponding
Bhattacharyya parameters can be defined.
\begin{align}
Q_{n} \rightarrow \left(Q_{1}^{n-1}, Y_{1}^{N/2}\right) \notag
\end{align}
\begin{align}
R_{n} \rightarrow \left(R_{1}^{n-1}, Y_{N/2+1}^{N}\right) \notag
\end{align}
\begin{align}
Z(Q_n) = \sum_{q_1^{n-1}, y_1^{N/2}} \sqrt{{\mathbb
P}(q_1^{n-1},y_1^{N/2}|0){\mathbb P}(q_1^{n-1},y_1^{N/2}|1)} \notag
\end{align}
\begin{align}
Z(R_n) = \sum_{r_1^{n-1}, y_{N/2+1}^{N}} \sqrt{{\mathbb
P}(r_1^{n-1},y_{N/2+1}^{N}|0){\mathbb P}(r_1^{n-1},y_{N/2+1}^{N}|1)}
\notag
\end{align}
In \cite{arikan09}, it is shown that for even-indexed $U_{2n}$,
\begin{align}
Z\left(U_{2n}\right)=Z\left(Q_n\right)Z\left(R_n\right)
\end{align}
and for odd-indexed $U_{2n-1}$,
\begin{align}
Z(U_{2n-1})\leq Z(Q_n) + Z(R_n) - Z(Q_n)Z(R_n)
\end{align}
Therefore, some upper bounds of $Z(U_n)$ can be recursively
calculated. This recursive process starts with upper bounding the
Bhattacharyya parameters for the $W_2$ channels based on $Z(X_n)$,
where $Z(X_n)$ is the Bhattacharyya parameter for the channel
$X_n\rightarrow Y_n$. The Bhattacharyya parameters $Z(U_n)$ are
important, because $\sum_{n\in {\mathcal M}}Z(U_n)$ is an important
indicator of the decoding reliability. In fact, it is shown in
\cite{arikan09} that $\sum_{n\in {\mathcal M}}Z(U_n)$ is an upper
bound of the block error probability.

In the sequel, we may abuse the notation and use $Z(U_n)$ to denote
the above recursively calculated upper bounds of the Bhattacharyya
parameters without introducing ambiguity. Because the exact values
of $Z(U_n)$ will not be calculated in this paper. Whenever $Z(U_n)$
denotes an actual number as in an algorithm, it denotes the above
recursively calculated upper bound.

\section{Virtual Equivalent Channel Method for Decoding Reliability Estimation}
\label{section_equ_compound_channel}

In this section, we present the proposed method for determining the
decoding reliability for each message bit based on the already
received channel observations. That is, if the transmit message is
decoded based on the already received channel observations, then how
reliable the decoding result is for each message bit. Such decoding
reliability information may be used for adapting the proposed coding
scheme. For example, the future transmit symbol is selected, so that
the decoding reliability can be most significantly improved. In this
paper, we focus our discussions on the Binary Input Additive White
Gaussian Noise (BIAWGN) channels. However, our methods can be
generalized to other channel models as well.

In the sequel, we assume that the feed-forward channel is BIAWGN,
where $Y = V + Z$, $V\in\{-1,+1\}$ is the transmitted signal, $Z$ is
the additive noise with variance $\sigma_n^2$ and $Y$ is the
received channel observation. We assume that a BPSK modulation is
used. That is, if $X_n=0$, then the signal $V=-1$ is transmitted,
and if $X_n=1$, then the signal $V=1$ is transmitted.

\begin{figure}[h]
 \centering
 \includegraphics[width=3in]{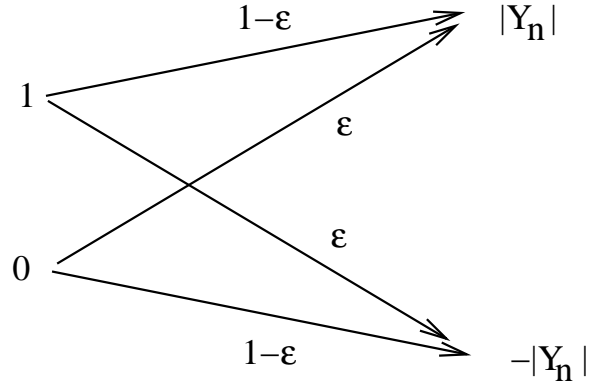}
 \caption{Equivalent BSC channel}
 \label{equivalent_bsc}
\end{figure}

The proposed decoding reliability estimation method is based on a
virtual equivalent channel approach. The key observation is that,
given the amplitude of the channel observation $|Y_n|$, the channel
between $X_n$ and $Y_n$ can be considered as a binary symmetric
channel, such that
\begin{align}
& {\mathbb P}(-|Y_n||X_n=0) = \alpha
\exp\left(\frac{-\left(|Y_n|-1\right)^2}{2\sigma_n^2}
\right) \notag \\
& {\mathbb P}(|Y_n||X_n=0) = \alpha
\exp\left(\frac{-\left(|Y_n|+1\right)^2}{2\sigma_n^2}
\right) \notag \\
& {\mathbb P}(-|Y_n||X_n=1) = \alpha
\exp\left(\frac{-\left(|Y_n|+1\right)^2}{2\sigma_n^2}
\right) \notag \\
& {\mathbb P}(|Y_n||X_n=1) = \alpha
\exp\left(\frac{-\left(|Y_n|-1\right)^2}{2\sigma_n^2} \right)
\label{virtual_bsc}
\end{align}
where $\alpha$ is a normalization constant. The equivalent binary
symmetric channel is shown in Fig. \ref{equivalent_bsc}, where the
channel parameter
\begin{align}
\epsilon = \frac{
\exp\left(\frac{-\left(|Y_n|+1\right)^2}{2\sigma_n^2} \right) }{
\exp\left(\frac{-\left(|Y_n|-1\right)^2}{2\sigma_n^2} \right) +
\exp\left(\frac{-\left(|Y_n|+1\right)^2}{2\sigma_n^2} \right) }
\end{align}

In other words, we consider the probability space of $X_n, Y_n$. We
form a filtration for this probability space, where the first
$\sigma$-algebra is generated by the random events $\{|Y_n|<c\}$,
and the second $\sigma$-algebra contains all the information about
$X_n$ and $Y_n$. Conditioned on the first $\sigma$-algebra and the
amplitude $|Y_n|$ of the received channel observation, we may
redefine the Bhattacharyya parameter $Z(X_n)$ of the channel between
$X_n$ and $Y_n$ as the Bhattacharyya parameter of the binary
symmetric channel in Fig. \ref{equivalent_bsc},
\begin{align}
Z(X_n) = \frac{
2\exp\left(\frac{-\left(|Y_n|^2+1\right)}{\sigma_n^2} \right) }{
\exp\left(\frac{-\left(|Y_n|-1\right)^2}{2\sigma_n^2} \right) +
\exp\left(\frac{-\left(|Y_n|+1\right)^2}{2\sigma_n^2} \right) }
\end{align}
This Bhattacharyya parameter $Z(X_n)$ measures the decoding
reliability for the symbol $X_n$ with the information of $|Y_n|$.
Clearly, the total reliability measure $\sum_{n\in {\mathcal M}}
Z(U_n)$ can also be updated with the updated $Z(X_n)$.

In the  case that one symbol $X_n$ is transmitted through the
feed-forward channel $K$ times and channel observations
$y_1,y_2,\ldots,y_K$ have been observed, we may consider a virtual
equivalent channel, where the input is binary  and the output
alphabet consists of $2^K$ symbols with the form
$\{Y_1,\ldots,Y_K\}$, $Y_k\in\{-y_k,y_k\}$. The corresponding
Bhattacharyya parameter $Z(X_n)$ is
\begin{align}
Z(X_n) = \prod_{k=1}^{K}\sum_{Y_k\in \{y_k,-y_k\}}\sqrt{{\mathbb
P}(Y_k|X_n=0){\mathbb P}(Y_k|X_n=1)}
\end{align}
There also exist other equivalent channel models for calculating the
Bhattacharyya parameter $Z(X_n)$.

\section{Polar Coding Scheme with Feedback}
\label{section_polar_with_feedback}

In this section, we present the proposed low-complexity polar coding
schemes using receiver feedback. We call such  schemes as the {\it
symbol-index-feedback} schemes. We discuss two versions of the
proposed schemes. The first version is called fixed-length
symbol-index-feedback scheme. For each block of $K$ transmit
information bits, the proposed coding scheme includes three stages.

{\bf Initial Stage:} encode the transmit $K$ information bits using
a conventional polar code with blocklength $N$ and rate $K/N$.
Transmit a predefined number of bits of the codeword $X_1^N$ using
the feed-forward channel.

{\bf Interactive Stage:} during each time slot, the receiver sends
one request of symbol using the feedback channel. The request of
symbol message contains the requested symbol index $n$, $1\leq n
\leq N$. After receiving the index $n$ of the request symbol, the
transmitter sends the request bit $X_n$ through the feed-forward
channel. After receiving the transmit bit, the receiver updates the
Bhattacharyya parameter $Z(X_n)$ and $\sum_{i\in {\mathcal M}}
Z(U_i)$ according to the rules in Section
\ref{section_equ_compound_channel}. The coding scheme may repeat the
above interactive process for a pre-designed fixed number of time,
and then goes to the stop stage. The symbol index $n$ should be
chosen, so that the decoding reliability can be significantly
improved. In this paper, we propose an approach of determining $n$,
such that
\begin{align}
n = \mbox{argmax}_j \frac{\partial \left[\sum_{i\in {\mathcal M}}
Z(U_i)\right]}{\partial \left[Z(X_j)\right]}\label{determine_n_eq}
\end{align}
That is, $\sum_{i\in {\mathcal M}} Z(U_i)$ descends most
significantly, if $Z(X_n)$ decreases.

{\bf Stop Stage:} a polar decoding algorithm, such as the decision
feedback decoding algorithm, is used. The scheme then outputs the
decoding results.

%%%%%%%%%%%%%%%%%%%%%%%%%%%%%%%%%%%%%%%%%%%%%%%%%%%%%%%%%%%%%%%%%%%%%%%%%%%%%%%%%%%%%%

The second version is called variable-length symbol-index-feedback
scheme using error-detection. For each block of $J$ transmit
information bits, the coding scheme includes three stages.

{\bf Initial Stage:} first encode the transmit $J$ information bits
using one conventional error-detection code into a codeword with
length $K$. Encode the $K$ error-detection codeword bits using one
conventional polar code with blocklength $N$ and rate $K/N$.
Transmit a predefined number of bits of the polar codeword $X_n$,
$1\leq n \leq N$ using the feed-forward channel.

{\bf Interactive Stage:} similarly as in the fixed-length scheme.
After the fixed number of bits are transmitted, the coding scheme
goes to the following stop stage.

{\bf Stop Stage:} a polar decoding algorithm, such as the decision
feedback decoding algorithm, is used. The scheme then outputs the
decoding results, if no decoding error is detected by the
error-detection code. Otherwise, the scheme returns to the
interactive stage.

%%%%%%%%%%%%%%%%%%%%%%%%%%%%%%%%%%%%%%%%%%%%%%%%%%%%%%%%%%%%%%%%%%%%%%%%%%%%%%%%%%%%%%%%

%

\section{Numerical Results and Discussions}

\label{sec_numerical}

\begin{figure}[h]
 \centering
 \includegraphics[width=3in]{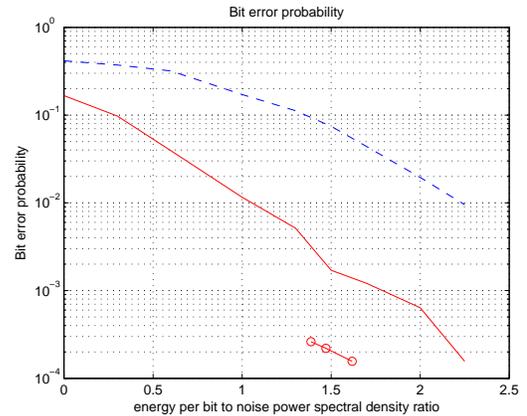}
 \caption{Bit error probabilities of the proposed coding schemes and the conventional polar codes.
 The bit error probabilities of the proposed fixed-length symbol-index-feedback scheme, the proposed
variable-length symbol-index-feedback scheme and
 the conventional polar code are shown by the solid line, solid line with circles, and dashed line respectively. }
 \label{bit_error_1024}
\end{figure}

\begin{figure}[h]
 \centering
 \includegraphics[width=3in]{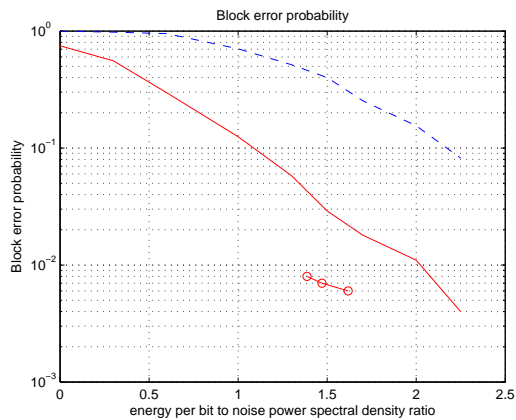}
 \caption{Block error probabilities of the proposed coding schemes and the conventional polar codes.
 The block error probabilities of the proposed fixed-length symbol-index-feedback scheme, the proposed variable-length symbol-index-feedback scheme and
 the conventional polar code are shown by the solid line, solid line with circles, and dashed line respectively.
 }
 \label{block_error_1024}
\end{figure}

In this section, we present simulation results for the proposed
symbol-index-feedback polar coding schemes. In Fig.
\ref{bit_error_1024}, we show the bit error probabilities of the
symbol-index-feedback schemes, and the conventional polar codes. The
solid line shows the bit error probabilities of the fixed-length
symbol-index-feedback polar coding scheme. The solid line with
circles shows the bit error probabilities of the variable-length
symbol-index-feedback polar coding scheme. The dashed line shows the
bit error probabilities for the conventional polar code without
using receiver feedback. The block error probabilities of the above
three coding schemes are shown in Fig. \ref{block_error_1024}, where
the block error probabilities of the fixed-length
symbol-index-feedback scheme, variable-length symbol-index-feedback
scheme, and the conventional polar code are shown by the solid line,
solid line with circles, and dashed line respectively. The $X$-axis
of the two figures shows the $E_b/N_0$, the energy per bit to noise
power spectral density ratio in dB as defined in \cite[Sec.
4.2]{forney_lecture}. All the above coding schemes use one baseline
polar code with blocklength $1024$ bits and rate $0.5$. The channel
is a AWGN channel with noise variance $1$. Each codeword bit is
transmitted with power $0.25$. The numbers of codeword bit
transmission (or average numbers) are determined by the $E_b/N_0$.

From the above figures and other simulation results that we have
obtained for many different cases of block lengths, coding rates
etc, we conclude that the symbol-index-feedback polar coding schemes
achieve significantly lower bit and block error probabilities
compared with the conventional polar codes. The performance
improvement is due to the fact that in the proposed schemes, the
data transmission processes are adapted to the already received
channel observations. The receivers are able to determine the part
of the transmit bits that can not be reliably decoded, and request
the codeword bits that can increase the decoding reliability most
significantly being transmitted. The data transmission processes are
therefore steered to arrive at reliable decoding results.

The variable-length version of the proposed schemes outperforms the
fixed-length version, This performance improvement is due to
flexible energy allocation between transmit bit blocks. The
variable-length scheme allocates less power for the blocks that can
be reliably decoded. The saved power may be used for some other
blocks, where the noise is atypical and decoding errors are more
likely. In fact, similar phenomenon that variable-length schemes
have better performance have been observed in the previous research,
see for example, \cite{polyanskiy10}, \cite{como09}, \cite{Sahai05}
etc.

The proposed coding schemes are different from many previous coding
schemes using feedback. In the proposed schemes, the requested
symbol indexes are feedback. While, many previous coding schemes
transmit the quantized channel observations back to the transmitter.
One main advantage of the proposed coding schemes is that almost all
computational complexities are at the receiver side. The proposed
coding schemes are favorable choices for the communication
scenarios, where the transmitters are power and complexity
constrained. Such low power and complexity devices may include many
embedded devices in ubiquitous computing, RFIDs, embedded sensors in
sensor networks etc.

\section{Conclusion}

\label{sec_conclusion}

In this paper, we propose novel error-control coding schemes using
receiver feedback based on polar codes. Two variations of the
proposed {\it symbol-index-feedback} schemes are discussed,
including one fixed-length scheme and one variable-length scheme. By
simulation results, we show that the proposed coding schemes achieve
significantly lower error probabilities compared with the
conventional polar codes. The variable-length scheme outperforms the
fixed-length scheme in terms of error probabilities. The proposed
symbol-index-feedback schemes have very low computational
complexities at the transmitter side. The proposed schemes are
favorable choices for communication scenarios, where the
transmitters are power and complexity constrained, such as embedded
devices in ubiquitous computing, wireless sensors, and RFID tags
etc.

%\nocite{*}
\bibliographystyle{IEEEtran}
\bibliography{the_bib}

\end{document}